\documentclass[useAMS,usenatbib]{mn2e}
\usepackage{graphicx,mathtools}
\usepackage{amssymb}
\usepackage[dvipsnames]{xcolor}
\usepackage{xparse,xcolor}
\usepackage{xspace}
\usepackage{cooltooltips}

\newcommand{\mnras}{MNRAS}

\title[No repulsive force in General Relativity]
{No repulsive force in General Relativity}
\author[M.A. Abramowicz \& J.-P. Lasota]{M. A. Abramowicz$^{1,2}$\thanks{E-mail:
marek.abramowicz@physics.gu.se} and J.-P.  Lasota$^{1,3}$\\
$^{1}$ Nicolaus Copernicus Astronomical Center, Polish Academy of Sciences, Bartycka 18, 00-716 Warszawa,
Poland\\
$^{2}$ G\"oteborg University, Physics Department, SE-412-96 G\"oteborg, Sweden\\
$^{3}$ Institut d'Astrophysique de Paris, CNRS et Sorbonne Universit\'es, UPMC Univ Paris 06, UMR 7095, 98bis Bd Arago, 75014 Paris, France\\
}

\begin{document}

\date{Accepted ---. Received ---; in original form ---}

\pagerange{\pageref{firstpage}--\pageref{lastpage}} \pubyear{2015}

\maketitle

\label{firstpage}

\begin{abstract}
We show that a recent assertion that gravitational wave emission can lead to a repulsive force explaining the accelerated expansion of the Universe is totally unfounded.
\end{abstract}


\section{Introduction}
\label{sec_intro}

In a recent paper \citet{GV16} asserted that ``reduction of the gravitational mass of the system due to emitting gravitational waves leads to a repulsive gravitational force'' and that in the cosmological context this implies that ``mergers of inhomogeneities like black
holes, resulting in emission of gravitational waves, can generate a repulsive gravitational force (...). These mergers act as
effective dark energy, if the total mass of the universe is decreased''. They conclude that ``This may imply that big bang and
accelerated expansion of the Universe is not related to current processes in the Universe but to a relict repulsive gravitational
force or to a configuration of space-time that originates in the previous cycle of the Universe when at the last stage of a
collapse the intensive generation of gravitational waves resulted in a sharp decrease of the gravitational mass of the Universe
(...).'' In this note we will show that these assertions are mistaken: there is no repulsive force in General Relativity, and the effect the authors have in mind cannot contribute to the accelerated expansion of the Universe.

\section{No repulsion and no acceleration}

It is obvious that what \citet{GV16} call a ``repulsive force" is simply the result of decreasing {\sl Newtonian} gravitational attraction when gravitating bodies loose mass. From this trivial fact \citet{GV16} drew cosmic conclusions.  They used a model of the Universe which is thought of as a finite sphere; radiation may leave it so its ``total mass'' decreases. The model, despite a relativistic make-up\footnote{Equation (11) of their paper contains a printing error: in the denominator of the first LHS term there should be $g_{00}$ instead of $g_{11}$.}, is basically Newtonian. This is why terms that are meaningless in Einstein's theory, such as ``the (cosmological) gravitational force'' or ``the total mass of the Universe'', are used. 

\citet{GV16} consider gravitational waves as the only agent that decreases the mass of radiating objects; they do not explain why
they neglect electromagnetic radiation -- after all, stars radiate, losing gravitational mass, which (by the authors' logic)
should also lead to a ``decrease of the total mass of the Universe'' and ``repulsive gravitational forces''. The electromagnetic
radiation effect  should be (by the authors' own logic) even stronger than the gravitational wave (GW) effect they discuss,
because the cosmological parameter $\Omega_{GW} \sim  10^{-9}$ for gravitational waves \citep{A1} is more than two orders of magnitude smaller than that corresponding to infrared background radiation \citep{HD01}.

Of course, the Universe is not a finite Newtonian sphere. Radiation (electromagnetic, gravitational) which leaves one particular
comoving volume does not disappear from the Universe -- it enters its other parts. The total density $\rho$ and the total pressure
$p$ are sums of several components, and the gravitational waves contribute to them. Neither $\rho_{GW}$ nor $p_{GW}$ are
negative. Therefore, as the Friedmann equation shows,
\begin{equation}
\frac{\ddot a}{a} = -\frac{4\pi G}{3}\left( \rho + \frac{3p}{c^2} \right),
\end{equation}
gravitational waves {\it always} contribute to {\it deceleration} of the expansion of the cosmological scale factor $a(t)$.

\section{Conclusion}
There is no gravitational repulsive force in General Relativity so it cannot contribute to the accelerated expansion of the Universe.

\section*{Acknowledgments}

This work was supported in part by the Polish National Research Center (NCN) grant No. 2015/19/B/ST9/01099.
JPL was supported by a grant from the French Space Agency CNES.  


\begin{thebibliography}{}

\bibitem[Abbott et al.(2016)]{A1} Abbott, B.~P., Abbott, R., Abbott, T.~D., et al.\ 2016, Phys. Rev. Letters, 116, 131102 

\bibitem[Gorkavyi \& Vasilkov(2016)]{GV16} Gorkavyi, N., \& Vasilkov, A.\ 2016, \mnras, 461, 2929 

\bibitem[Hauser \& Dwek(2001)]{HD01} Hauser, M.~G., \& Dwek, E.\ 2001, ARA\&A, 39, 249 



\end{thebibliography}

\label{lastpage}

\end{document}